\documentclass[12pt,preprint]{aastex}
\usepackage{lscape}
\usepackage{amsmath}

\def\day{{\,\rm day}}
\def\yr{{\,\rm yr}}

\def\d{{\,\rm d}}
\def\hr{{\,\rm hr}}

\def\EE{\rm eccentric envelope}
\def\EEP{\EE{} population\,}
\def\EB{\rm EB}
\def\circ{\rm circ}

\begin{document}
\title{Exploring a Stream of Highly-Eccentric Binaries with {\it Kepler}}
\author{Subo Dong\altaffilmark{1,2}, Boaz Katz\altaffilmark{1,3,4} 
and Aristotle Socrates\altaffilmark{1,3}}

\altaffiltext{1}{Institute for Advanced Study, Princeton, NJ 08540, USA}
\altaffiltext{2}{Sagan Fellow}
\altaffiltext{3}{John Bahcall Fellow}
\altaffiltext{4}{Einstein Fellow}

\begin{abstract}
With 16-month {\it Kepler} data, $14$ long-period ($40\d-265\d$) 
eclipsing binaries on highly eccentric orbits (minimum $e$ between 
$0.5$ and $0.85$) are recognized from their closely separated 
primary and secondary eclipses ($\Delta t_{\rm I,II}= 3\d-10\d)$. 
These systems confirm the existence of a previously hinted binary 
population situated near a constant angular momentum track at 
$P(1-e^2)^{3/2}\sim 15\d$, close to the tidal circularization period $P_{\circ}$. 
They may be presently migrating due to 
tidal dissipation and form a steady-state stream ($\sim 1\%$ of 
stars) feeding the close-binary population (few-$\%$ of stars). If so, future 
{\it Kepler} data releases will 
reveal a growing number (dozens) of systems at longer periods, 
following $\d N/\d \lg P \propto P^{1/3}$ with increasing eccentricities  
reaching $e\rightarrow 0.98$ for $P\rightarrow 1000\d$. 
Radial-velocity follow up of long-period eclipsing binaries with 
no secondary eclipses could offer a significantly larger sample.
Orders of magnitude more (hundreds) may reveal their presence 
from periodic ``eccentricity pulses'', such as tidal ellipsoidal variations, 
near pericenter passages. Several new few-day-long eccentricity-pulse candidates 
with long period (P=$25\d-80\d$) are reported.
\end{abstract}

\section{Introduction}
\label{sec:intro}
How do close binaries $(P\lesssim10\d)$ form? An intriguing 
clue comes from their surprisingly high fraction having a 
tertiary companion, reaching as much as $\sim 96\%$ for systems with
$P \lesssim 3\d$ \citep{tol06}. This lends support to the
``Kozai Cycles plus Tidal Friction'' (KCTF) formation channel advocated by several authors
\citep{egg98,egg01,fab07}. In KCTF, a tertiary companion on a  
highly inclined outer orbit can excite the inner binary orbit, initially 
having a long period, to very high eccentricity during Kozai-Lidov cycles
\footnote{Recent studies on Kozai-Lidov cycles show that the parameter 
space that allows excitation to very high eccentricity is substantially wider 
than previously thought due to the commonly neglected octupole contribution to the tertiary potential \citep{for00, nao11, kat11, lit11}.}, which 
allows efficient tidal dissipation to take place during close 
pericenter passages, eventually shrinking the orbital period to a few days. 

One clear implication of KCTF is that there should exist a population 
of highly-eccentric binaries in the process of migrating to 
their final close-in orbits. Such a prediction was studied in the context of 
a similar formation process of ``hot Jupiters'' (close-in Jovian-mass companions) 
by \cite{soc11}. In this Letter, we explore observational implications of such a population which is possibly responsible for close-binary formation.

\section{Steady-State Distribution}
\label{sec:steadystate}
In this section, we briefly review the features of a steady-state distribution of 
migrating binaries due to tidal dissipation in analogy to migrating Jupiters discussed in 
\citet{soc11}. Stars have been forming at a roughly constant rate over the age of 
the Galaxy, therefore one would expect a steady-state source of new distant 
binaries that constantly migrate to lower periods, ``feeding''
the close-binary population. Tidal dissipation drains orbital energy but 
hardly changes the angular momentum of an eccentric binary. While losing energy 
due to tidal dissipation, eccentric binaries migrate on tracks of constant 
orbital angular momentum corresponding to constant $P(1-e^2)^{3/2} = P_f$.  
$P_f$ equals the period of the final circularized close binary (once circular, tidal dissipation ceases to operate in the tidally locked system). The rate of energy dissipation is a strong function of binary separation, and for an eccentric orbit, most of the energy loss occurs during the pericenter passage.
At high $e$, due to the constant angular momentum, the pericenter passages have approximately the 
same parabolic shape, and the energy lost each orbit, $\Delta E$, is approximately the same. This implies an average energy loss rate $\dot{E} \propto 1/P$, which results in a steady state distribution $\d{N}/\d{\lg P} \propto E/\dot{E}\propto P^{1/3}$, where $E\propto P^{-2/3}$.
This distribution at high $e$ is relatively independent of tidal 
dissipation physics.  At moderate and low $e$, the distribution is sensitive to the poorly understood
tidal dissipation processes. The distribution obtained for the commonly assumed 
equilibrium tide model (e.g, \citealt{hut81}) is derived in \citet{soc11}. An 
additional complication is that angular momentum oscillates while the Kozai 
mechanism operates. As shown in  \citet{soc11}, this does not significantly 
affect the resulting distribution as long as we consider narrow bins 
in angular momentum (or $P_f$) in which most of the dissipation occurs.  In this Letter, we focus on the long-period, highly eccentric members in this population, which are simple to analyze theoretically and, as shown below, are easily distinguishable observationally.

\section{Past Observational Hints} 
\label{sec:hint}
Making meaningful statistical studies of the binary distribution requires an un-biased 
sample. The commonly adopted study on multiplicity of solar-type primaries
is from a complete, volume-limited (within 22 pc and spectral type F7-G9) sample of $\sim 164$ 
systems by \citet{may91}. They identify three period regimes. Binaries with 
periods shorter than about $10\d$ have eccentricities close to zero and are thought  to be efficiently circularized by the strong tidal dissipation at small separations. Those with periods longer than $\sim 1000\d$ are found to be consistent with an Ambartsumian 
distribution, $dN/de=2e$, which is expected if energy conserving dynamical perturbations 
operate. In the intermediate period range $10\d<P<1000\d$, the distribution in $e$ is 
described as bell-shaped with an average of $\bar e \sim 0.3$, which has since been 
widely accepted as reflecting the initial distribution at binary formation. They 
note that the largest eccentricities in the $10\d <P<1000\d$ regime seem to be curiously 
populated by triple systems. They further note that just beyond the selection criterion of their complete sample, there exists some clear outliers to the eccentricity distribution, and
interestingly two of them, HD 137763 ($e=0.975$, $P=890\d$) and HD 114260 ($e=0.56$, $P=20.5\d$)  
are found to have additional companions.

\cite{rag10} provide a new volume-limited sample selected using Hipparcos
parallax (within 25 pc and stellar type F6-K3) including 
454 solar-type stars, a factor of $\sim 3$ larger in size than that of \citet{may91}. 
Figure 1 shows the period and eccentricity distribution 
of binaries in this sample. Systems with known companions (multiples) 
are shown in filled circles while those without are plotted in crosses. 
As pointed out by the authors, the upper envelope of eccentricity for 
systems with $10\d <P< 1000\d$ appears to mostly consist of multiples, indicative of Kozai 
oscillations. The five ``\EE'' systems, each with eccentricity larger than that of 
any binary with shorter period, have the following parameters: $(P,e,P_{\rm tertiary}) 
=$ $(890 \d, 0.975, 2.4\times10^4 \yr)$, $(88 \d, 0.81, 700 \yr)$, 
$(48 \d, 0.61, 100 \yr)$, $(24 \d, 0.41, 2000 \yr)$, $(21 \d, 0.25, 260 \yr)$
\footnote{Orbital parameters taken from the 20 Apr 2010 updated version of 
\citet{msc97}}.

These envelope multiples, which possibly consist of $\sim 1\%$ of all main-sequence stellar systems, 
are very interesting in the context of KCTF. Figure 1 shows constant angular momentum 
tracks, as discussed in section ~\ref{sec:steadystate}, 
of $P_f = P(1-e^2)^{3/2} = 10\day\,{\rm and}\,30\day$, which enclose the envelope. 
This raises the exciting possibility that these systems may represent a population of 
highly eccentric binaries in steady-state KCTF migration. If true, the fact that the
$P_f$ values of these systems are close to the circularization threshold of $P_{\circ}  \sim 10\d$ is 
not surprising. Systems with $P_f \ll P_{\circ}$, circularize quickly, while those with  $P_f \gg P_{\circ}$, do not significantly migrate. Systems at $P_f \sim P_{\circ}$ migrate 
on a time scale comparable to the age of the Galaxy, implying that a significant fraction of them should be currently migrating.
 
Curiously, the tertiary companions of the above five systems all share very long, 
$P_{\rm tertiary} \sim 100 - 10^4 \yr$, orbital periods. 
More detailed examination of their orbital parameters indicate that due to 
the large perturber separations, the amplitude of the Kozai oscillations (if present) is suppressed by relativistic precession
(see the discussions on ``quenched'' Kozai oscillation in \citealt{soc11} and 
Figure 1 of \citealt{fab07} for an example), which would naturally explain why they form an upper eccentric envelope. 

Unfortunately, the small-number ($\sim 5$) statistics
makes it difficult to robustly establish the existence of this interesting \EEP{} and to investigate its properties.

\section{{\it Kepler} Probes}
\label{sec:kepler}

{\it Kepler} mission has great potential to provide a complete sample that is 
orders-of-magnitude larger than \cite{rag10} to probe the above-mentioned 
population between $20\d$ and $1000\d$ that consists of possibly $\sim 1\%$ of main-sequence stellar systems. 
{\it Kepler} is a high-cadence, high-precision mission that aims to find habitable Earth-size planets. It has been monitoring $\sim 150,000$ FGK stars continuously with high cadence and high photometric precision since 2009, and is designed to operate for $3.5 \yr$ with the possibility of an extension of several more years. The latest data public release available on 10 January, 2012 includes 16-month long light curves. 
 There are potentially $\sim 1500$ systems in total belonging to the \EEP among {\it Kepler} targets.

\subsection{Eclipsing Binaries}
\label{sec:eb}
 A straightforward method to detect binary stars is to look
for eclipses i.e., eclipsing binaries (EBs). For an isotropic distribution of orbital orientations, the chance of seeing an EB\, with at least one of the stars being eclipsed 
is ${\cal P}_{\rm EB} = R_{\rm tot}[a(1-e^2)]^{-1}(1+2e/\pi)$ if the sum of the radii of the two stars $R_{\rm tot}$ is  much smaller than their separation. This geometrical probability is roughly constant for a given orbital angular momentum $J\propto [a(1-e^2)]^{1/2}$, and for a 
track with $P_f = 15 \day$ and high $e$, is approximately ${\cal P}_{\rm EB} = 0.06 (R_{\rm tot}/R_\odot)(M_{\rm tot}/M_\odot)^{-1/3}$, where $M_{\rm tot}$ is the total binary mass. We therefore expect to find of order $\sim 100$ \EB s within the \EEP with $20\d <P < 1000 \d$. 

While the period is a direct observable of an \EB\, light curve, the eccentricity is not. One way to accurately determine $e$ is to make radial velocity observations. A second option is to carefully model the eclipse light curve using detailed characterization of the stellar parameters, allowing for a reasonable estimate of the eccentricity.  
In this letter we follow a third path by focusing on a subsample of \EB s with both primary and secondary eclipses observed, upon which one can place strict lower limits on the eccentricity by measuring the timing separation between the two eclipses.
We show that a robust selection of highly eccentric binaries can be made by searching for long period eclipses with short primary-secondary separations.    

The probability of observing both eclipses is given by ${\cal P}_{\rm EB,2} = (\pi-2e)/(\pi+2e) {\cal P}_{\rm EB}$, which amounts to about  $\sim 20\%$ of all eclipsing systems at high $e$. We thus expect to find about two dozen systems from the \EEP. The timing difference between the eclipses, $\Delta{t_{I,II}} = t_{\rm II}-t_{\rm I}$ is given by 
\begin{equation}
\Delta{t_{\rm I,II}} (P, e, \omega)= 
\frac{P}{\pi} \left[\arccos\left(\frac{e\cos\omega}{\sqrt{1-e^2\sin^2\omega}}\right) 
-\sqrt{1-e^2}\frac{e\cos\omega}{1-e^2\sin^2\omega}\right],
\end{equation}
where $\omega$ is the argument of pericenter. At high $e$ the eclipses are most likely to occur near pericenter  and the separation between the two eclipses is of order the pericenter passage time $\sim P(1-e)^{3/2}/2$. The separation reaches minimum at $\omega=0$ and is given by 
\begin{equation}\label{eq:Dtmin}
\Delta{t_{\rm I,II, min}}(P, e) = \frac{P}{\pi} (\arccos e - e\sqrt{1-e^2})
\xrightarrow[e\rightarrow1]{}\frac{2P_f}{3\pi}.
\end{equation}
and is approximated by $P(1-e)^{3/2}/2$ to an accuracy of better than $20\%$. 

\begin{figure}[h]
\epsscale{0.8} \plotone{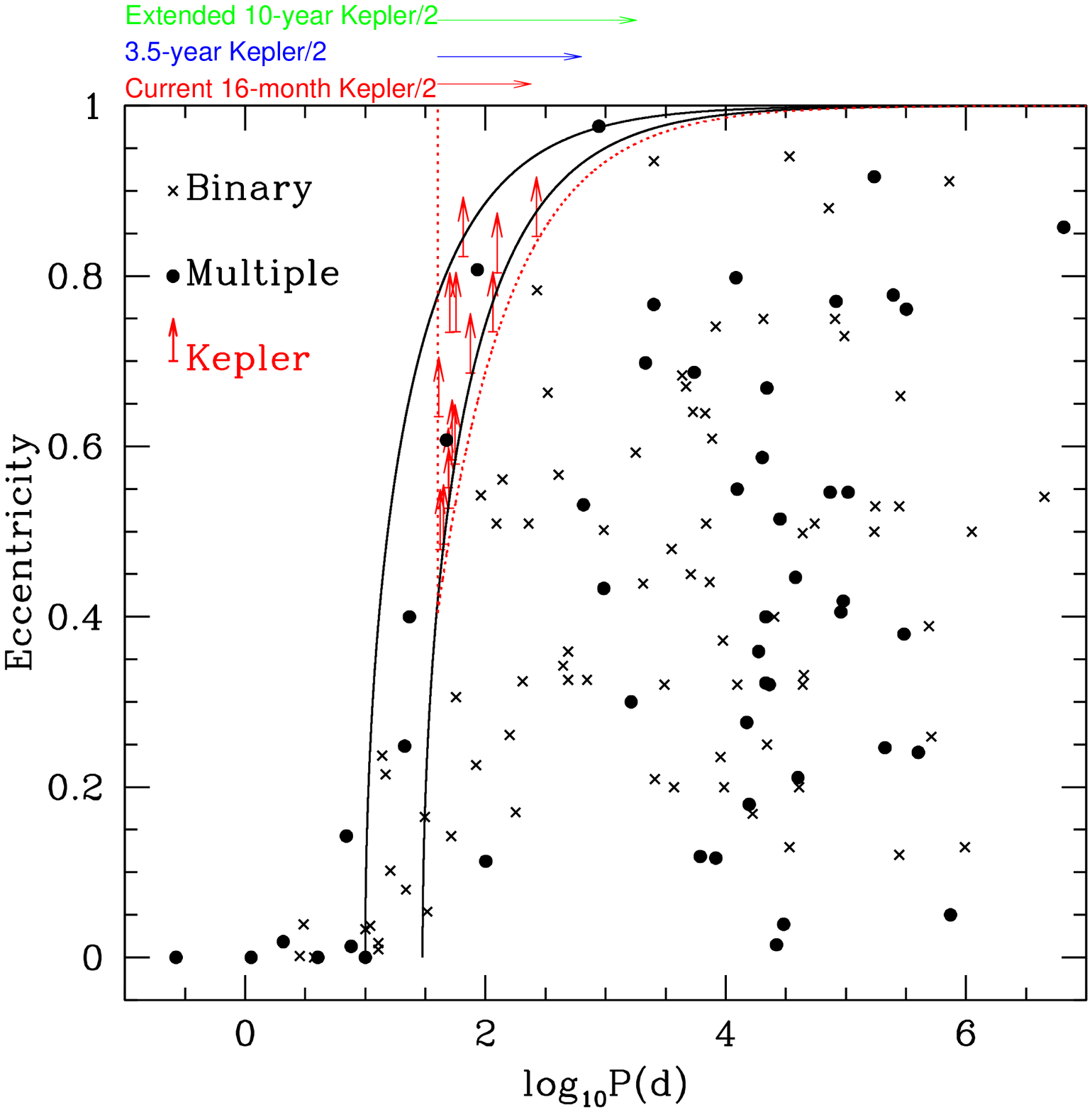}
\caption{Period and eccentricity of stellar systems 
presented in the \citet{rag10} volume-limited FGK dwarf 
multiplicity survey sample. Components within multiples are shown
as filled circles while the binaries with no detected additional 
companions as crosses. The upper envelope in eccentricity 
for $10 \day <P<1000 \day$ seems to be defined by a population 
of eccentric systems in multiples. They are located within a 
narrow range of constant angular momentum tracks  
at $P_f = P(1-e^2)^{3/2}$ between $10 \day {\,\rm and \,} 30 \day$, plotted as solid lines. 
The period and minimum eccentricity of the long-period, highly-eccentric 
eclipsing binary sample described in this paper are shown in red arrows.
The latter is robustly constrained by timing differences between primary
and secondary eclipses $t_{II} - t_{I}$. 
The dotted lines in red illustrate the observational cuts placed in 
searching for these eclipsing binaries. One is a lower cut in period 
of $P>40\d$, and the other corresponds to an upper cut in eclipse timing 
difference $t_{II} - t_{I} < 10\d$, shown by the eccentricity lower
limits that it sets as a function of period. Half of the duration for the current (
$\sim 16-$month), expected full (3.5 $\rm yr$) and possibly extended ($\sim 10\rm yr$)
 Kepler data releases are plotted as arrows, respectively, at the top of the figure, illustrating roughly the expected
completeness period threshold for eclipsing binary detections.
\label{fig:rag10}}
\end{figure}

We search for \EB s with periods longer than $P_{\rm cut} = 40\d$ with secondary eclipses separated by less than $\Delta{t_{\rm I,II,cut}}=10\d$ from the primary eclipses. Any \EB s that survive the latter cut must have an eccentricity larger than $e_{\rm cut}$, which satisfies $\Delta{t_{\rm I,II,\min}}(e_{\rm cut})=\Delta{t_{\rm I,II,cut}}$. The resulting minimal eccentricity
at $\Delta t_{\rm I,II,cut} =10\d$ as a function of $P$ is shown in a curved dashed line in Figure \ref{fig:rag10}. The period 
cut $P_{\rm cut} = 40\d$ is also shown as a dashed line. Any observed \EB{} that satisfies these observational cuts must reside in the ($P,e$) domain defined by these two lines, which allows for a robust selection of the \EEP. Moreover, $\gtrsim 80\%$ of systems with $P_f  = 15\d$ and $P>P_{\rm cut}=40 \d$ that show two eclipses, satisfy $\Delta t_{\rm I,II} < \Delta t_{\rm I,II,cut} = 10 \d$ implying that this cut does not significantly affect systems in the \EEP.

\begin{figure}[h]
\epsscale{1} \plotone{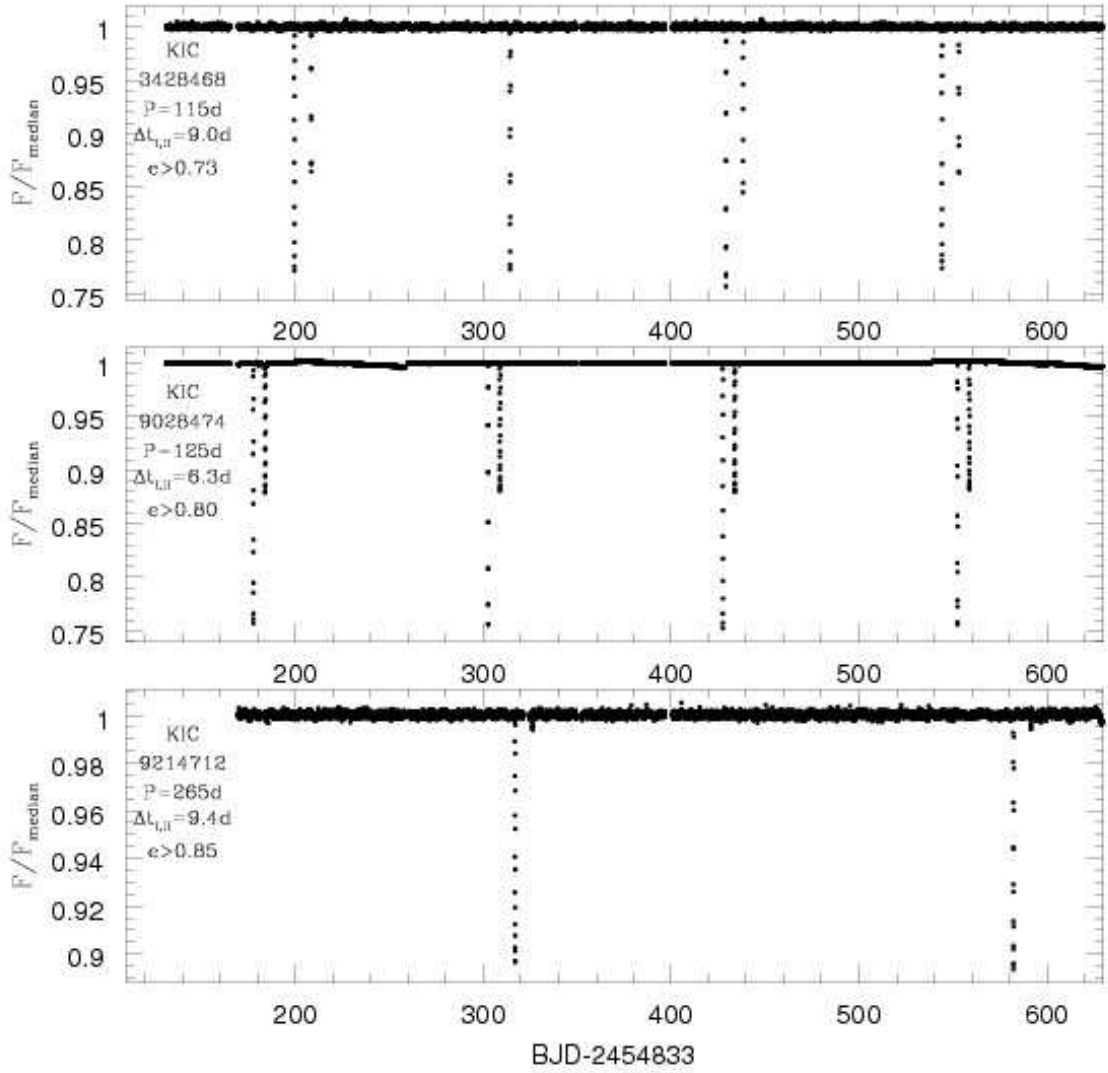}
\caption{Light curves for the three longest-period eclipsing binaries in our sample. Primary and secondary eclipses
can be readily recognized. The timing difference between the two eclipses place a 
robust lower limit to the orbital eccentricity, as discussed in section \ref{sec:eb}. \label{fig:lc}}
\end{figure}

We search for \EB s in the Janurary 2012 release of Kepler data, with all long-cadence ($29.4{\rm min}$),  Pre-Search data Conditioning (PDC) light curves from quarters Q1-Q6 (spanning $16$ months in total). First we identify eclipse candidates using a simple criterion that at least 3 consecutive points are above $3 \sigma$ deviations from the local median (determined with a window of 200 points). 
Then candidate light curves are selected by requiring pairs of candidate eclipses to repeat with periods longer than $40\d$ and that at least one eclipse be deeper than $1\%$. By visually inspecting each of the selected light curves, we identified 13 binary systems with double eclipses having timing separations smaller than 
$10\d$ (and periods greater than $40\d$). Six out of these are listed in the {\it Kepler} EB\, catalog \citep{sla11}, which is based on the {\it Kepler} Q0-Q2 data (with a total  time span of $125\day$). Among the binaries in the catalog, we found an additional system that survives our period and eclipse separation cuts, but  for which both eclipses have depths smaller than $1\%$. Therefore our selection appears to be efficient. The parameters of the 14 systems are listed in Table 1 and plotted as red arrows in Figure \ref{fig:rag10} showing 
their periods and minimal eccentricities derived from Equation \eqref{eq:Dtmin}, based on their measured $\Delta{t_{\rm I,II}}$.  The light curves of the three systems with the longest periods are shown in Figure \ref{fig:lc} for illustration. Clearly the search is incomplete at long periods, $P\gtrsim 160\d$, at which fewer than 3 primary eclipses occur during the observation time span. Our detections establish that at least $\sim 1\%$ of stellar systems are binaries (with additional possible companions) in the selected region in the ($P,e$) plane.

\subsection{Eccentric Ellipsoidal Pulses}\label{sec:epulses}
{\it Kepler}'s exquisite photometric precision opens up a venue of binary star 
detections with ellipsoidal, reflection/irradiation, and
relativistic beaming modulations 
(e.g. \citealt{loe03,fai11}). For very eccentric binaries, these 
modulations show up as short ``pulses'' that reach
maximum amplitude near pericenter and last about the pericenter 
passage time. \citet{koi54} reported the first {\it Kepler} detection of such 
``eccentric pulses'' in KOI-54, with $P=41.8\d$ and $e=0.83$. 
\citet{tho11} reported discoveries of 17 systems with $P<20\d$, 
dubbed as ``heartbeat stars''.

In the case that ellipsoidal variations dominate (for
small pericenter separation $r_p$), orbital parameters including
eccentricity, argument of pericenter and the inclination can be
robustly extracted from the shape of the pulses 
\citep{kum95}. In contrast to
\EB s, the eccentric ellipsoidal
variations can be virtually seen at all inclinations with similar
amplitude, thus in principle allowing the detection of orders of magnitude
more such systems than from \EB s. The amplitude of the ellipsoidal pulse due to a Sun-like
primary in a long-period, highly eccentric system with mass ratio $q$ and
$P_f \sim 15\d$ is roughly $\sim q (R_\odot/r_p)^{3} \sim 5\times10^{-4} q/(1+q)$
with a width of $\sim 3 \d$. \citet{jen10} demonstrates that
at {\it Kepler} magnitude $m_{\rm Kep}<14$ (half of all {\it Kepler} targets), the
majority of the {\it Kepler} dwarfs reach better than $\sim 10^{-4}$ precision
over $6.5\hr$ interval, including photometric noise and stellar variability.
Therefore the prospect of detecting a large number of Sun-like binaries
exhibiting eccentric ellipsoidal variations is promising.

\begin{figure}[h]
\epsscale{1} \plotone{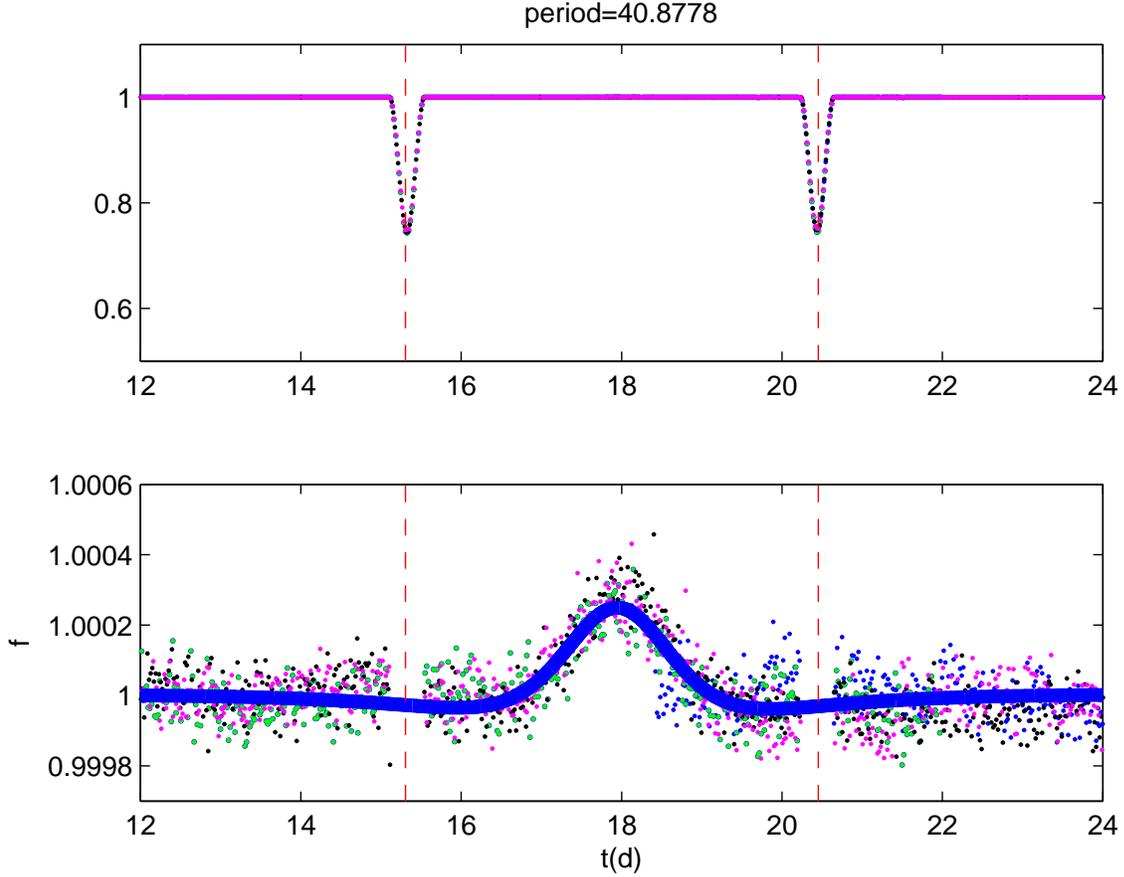}
\caption{Folded light curves of KIC 6864859, a 41-d eclipsing binary that exhibits ``eccentricity pulses'' with $\sim 2.5\times 10^{-4}$ amplitudes between the two eclipses. Points with different colors represent 
observations in different period folds. The upper panel shows the portion of the light curve with two eclipses. The lower panel plots the same portion of the light curve with a zoomed view to
show the eccentric pulses which are well described by the linear-perturbation model of ellipsoidal variations \citep{kum95} plotted in 
solid blue line with $e = 0.63$. The dashed lines indicate the eclipse timings required by the 
ellipsoidal variation model, which match well with the observed values.
 \label{fig:pulse}}
\end{figure}

\begin{figure}[h]
\epsscale{1} \plotone{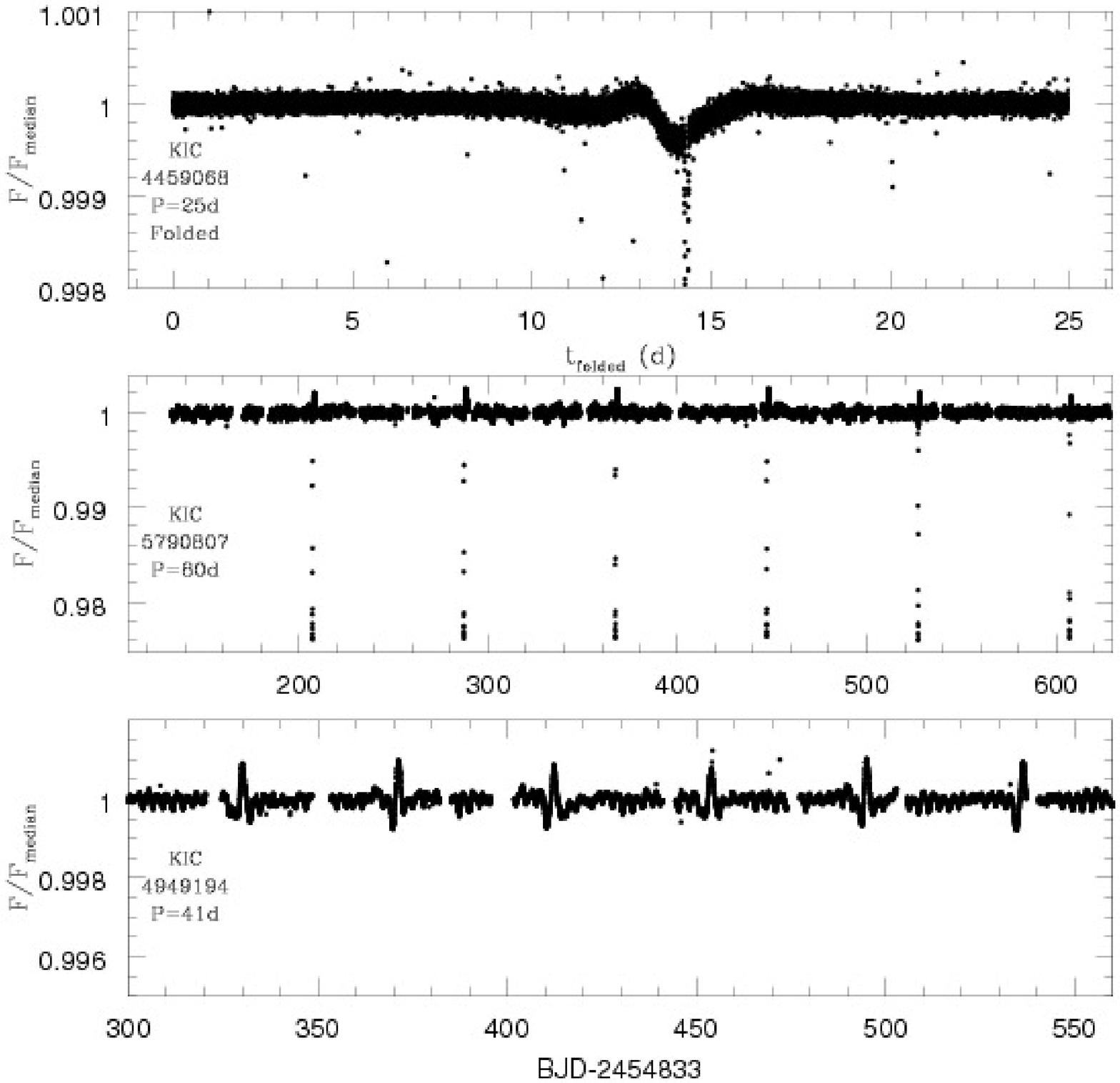}
\caption{Examples of systems exhibiting eccentricity pulses (see section \ref{sec:epulses}). The upper panel shows a folded light curve (KIC 4459068, P=25 d) while the light curves shown in the middle and 
lower panels are not folded (KIC 5790807, P=80d and KIC 4949194, P=41d). 
 \label{fig:pulsesExamples}}
\end{figure}

One simple way to find eccentric pulses is to examine the \EB{} light curves 
(with either one or both of the primary and secondary eclipses) and
inspect the region in the vicinity of the eclipses where such pulses
are most likely to occur. The known periodicity helps distinguishing
the pulses from other sources of variabilities. Figure \ref{fig:pulse} shows
the folded light curve of the \EB\, KIC $6864859$ with detection of 
pulses with amplitude of $0.00025 $. The pulse shape, shown in the bottom panel,
is well described by the linear-perturbation model of ellipsoidal variations
$\delta F \propto {[1-3\sin^2 i\cos^2(f+\omega)]}/r^3$
\citep{kum95}, where $f$ is the true anomaly and $r$ is the instantaneous binary separation,
for parameters ($e = 0.63$, $\omega = 2^{o}$ and $i = 90^{o}$), plotted in solid 
line. These parameters also agree with the timings of the primary and secondary eclipses, 
shown in dash lines. Clearly a pulse similar to that of KIC $6864859$ can be 
detected without the aid of eclipses. Since the geometric 
bias can be well estimated, one could make consistency checks between the systems
detected among the eclipsing binaries and the ones without the eclipses.
The light curves of a few other candidates (KIC $4459068$, $5790807$, $4949194$) 
with periods ranging from $25\d$ to $80\d$ are shown in Figure \ref{fig:pulsesExamples}. 
KIC $4459068$ and KIC $5790807$ are associated with single eclipses (both of which
are included in the \EB catalog by \citet{sla11}) while 
KIC $4949194$ does not exhibit any eclipses. Some other probable candidates include
KIC $9172506$ and $9028474$, which show possible pulses with amplitudes similar to 
those of other variabilities in the light curve. 
Note that the eccentricity pulse signals have relatively large 
amplitude for early-type stars, raising the interesting possibility of studying the possible 
eccentric migrating binaries for those stars.

\section{Discussion}\label{sec:Summary}
The presence of more than a dozen highly-eccentric, long-period {\it Kepler} 
eclipsing binaries confirm that $\sim 1\%$ of stellar systems 
reside  in the P-e plane near $P(1-e^2)^{3/2} = P_f \sim 15\d$, as hinted
by the \citet{rag10} sample. The binaries in this stream may be excited 
to high eccentricity in Kozai cycles and experience tidal migration in 
the process of forming close binaries. This interpretation has multiple 
implications that can be tested in the near future: (1) A growing population
at increasing periods is expected from the steady-state distribution, following 
$\d N/\d \lg P \propto P^{1/3}$. This can be probed by future releases
of {\it Kepler} data, which will enable the exploration of longer-period regimes. One interesting possibility is the identification of the location from which these binaries originate.  Larger samples can be
obtained by using Radial-Velocity followup of \EB s or by searching for eccentricity 
pulses in the {\it Kepler} data. (2) If affected by Kozai, these systems should have additional companions which can be searched for  
using high-resolution imaging. (3) Direct tests of Kozai oscillations can be achieved by preforming differential astrometry measurements of 
mutual inclinations between the inner binary and the companion (e.g., \citet{mut06}) 
for the nearby multiples in the stream.

\acknowledgements We thank Scott Tremaine, Tsvi Mazeh, Andy Gould, Jose Prieto and Zheng Zheng for useful discussions. Work by S.D. was performed under contract with the California Institute of Technology (Caltech) funded by NASA through the Sagan Fellowship Program. This research was partially supported by Minerva, ISF, and the Universities Planning \& Budgeting Committee grants. B.K is supported by NASA through Einstein Postdoctoral Fellowship awarded by the Chandra X-ray Center, which is operated by the Smithsonian Astrophysical Observatory for NASA under contract NAS8-03060.

\bibliographystyle{apj}

\begin{deluxetable}{cccc}
\tablewidth{0pt}
\tablehead{
\colhead{KIC}&\colhead{P(day)}&\colhead{$\Delta{t_{\rm I,II}}$(day)}&\colhead{$e_{\min}$}}
\startdata
9214712&265&9.4&0.85\\
9028474&125&6.3&0.80\\
3428468&115&9.0&0.73\\
5535280\tablenotemark{*}&75&7.5&0.69\\
11249624&66&2.9&0.82\\
3102000&57&4.5&0.73\\
10798605&56&8.6&0.58\\
2708614&53&7.9&0.58\\
9172506\tablenotemark{*}&50&4.0&0.73\\
2442084\tablenotemark{*}&50&9.0&0.53\\
5035972\tablenotemark{*\dag}&49&8.3&0.55\\
7541502\tablenotemark{*}&45&9.1&0.48\\
8553907\tablenotemark{*}&42&8.7&0.48\\
6864859\tablenotemark{*}&41&5.1&0.63\\\enddata
\tablenotetext{*}{Reported in \citep{sla11} }
\tablenotetext{\dag}{Not identified by our independent search (due to eclipse depth $<0.01$)}
\end{deluxetable}
\end{document}